\documentclass[reprint,aps,prd,twocolumn,notitlepage,nofootinbib,showpacs,preprintnumbers,superscriptaddress]{revtex4-2}
\usepackage[T1]{fontenc}
\usepackage[utf8]{inputenc}
\usepackage{bm}
\usepackage{amsmath}
\usepackage{amssymb}
\usepackage{mathrsfs}
\usepackage{esint}
\usepackage{color}
\usepackage{graphicx}
\PassOptionsToPackage{normalem}{ulem}
\usepackage{ulem}

\usepackage{times}
\usepackage{hyperref}

\begin{document}
\preprint{YITP-20-138, IPMU20-0112}

\title{Minimally modified gravity with an auxiliary constraint: a Hamiltonian
construction}

\author{Zhi-Bang Yao}%
      \email[Email: ]{yaozhb@mail2.sysu.edu.cn}
      \affiliation{School of Physics and Astronomy, Sun Yat-sen University, Guangzhou 510275, China}

\author{Michele Oliosi}%
\email[Email: ]{michele.oliosi@yukawa.kyoto-u.ac.jp}   
\affiliation{Center for Gravitational Physics, Yukawa Institute for Theoretical
Physics,~\\
 Kyoto University, Kyoto 606-8502, Japan}     

\author{Xian Gao}%
    \email[Email: ]{gaoxian@mail.sysu.edu.cn}
    \affiliation{School of Physics and Astronomy, Sun Yat-sen University, Guangzhou 510275, China}

\author{Shinji Mukohyama}%
    \email[Email: ]{shinji.mukohyama@yukawa.kyoto-u.ac.jp
    }
\affiliation{Center for Gravitational Physics, Yukawa Institute for Theoretical
Physics,~\\
 Kyoto University, Kyoto 606-8502, Japan}
\affiliation{Kavli Institute for the Physics and Mathematics of the Universe (WPI),
~\\
The University of Tokyo Institutes for Advanced Study,~\\
The University of Tokyo, Kashiwa, Chiba 277-8583, Japan}

\date{November 2, 2020}

\begin{abstract}
Working directly with a general Hamiltonian for the spacetime metric with the $3+1$ decomposition and keeping only the spatial covariance, we investigate the possibility of reducing the number of degrees of freedom by introducing an auxiliary constraint. The auxiliary constraint is considered as part of the definition of the theory. Through a general Hamiltonian analysis, we find the conditions for the Hamiltonian as well as for the auxiliary constraint, under which the  theory  propagates two tensorial degrees of freedom only. The class of theories satisfying these conditions can be viewed as a new construction for the type-II minimally modified gravity theories, which propagate the same degrees of freedom of but are not equivalent to general relativity in the vacuum. We also illustrate our formalism by a concrete example, and derive the dispersion relation for the gravitational waves, which can be constrained by observations.
\end{abstract}

\maketitle

\section{Introduction}

To date, Einstein's General Relativity (GR) strongly stands against myriad tests, though this does not retract the necessity of inspecting its validity in new regimes and proposing alternative models of gravity (see e.g.~\cite{Clifton:2011jh}), i.e., modified gravity (MG). Notably, GR as well as some of the MG theories have had to pass the test of the most accurate detections ever\textemdash the gravitational wave (GW) signals generated by the mergers between black holes and/or neutron stars and detected by LIGO and Virgo \cite{Abbott:2016blz,TheLIGOScientific:2017qsa} and the joint arrival of the photons and gravitational waves emitted by one of the mergers \cite{Monitor:2017mdv}. Not only the information coming from the speed of the GW has already constrained several models of MG (see e.g.~\cite{Sakstein:2017xjx,Creminelli:2017sry,Baker:2017hug,Arai:2017hxj,Heisenberg:2017qka,Langlois:2017dyl,Ezquiaga:2017ekz}) but also the polarization information will be another important discriminator between GR and MG models by future GW measurements \cite{Abbott:2018utx,OBeirne:2019lwp}. Indeed, GR propagates two tensorial polarization modes only, i.e., two tensorial degrees of freedom (TTDOF), due to the spacetime diffeomorphism symmetry. Furthermore, GR is the unique 4-dimensional theory for the spacetime metric with the second order equations of motion, which obeys the general covariance and locality, as stated by Lovelock's theorem \cite{Lovelock:1970,Lovelock:1972vz}. A straightforward corollary is that GR is the unique theory propagating TTDOF if all the assumptions of Lovelock's theorem are preserved. 

However, what if we abandon at least one of the conditions of Lovelock's theorem: is there any space of theories for us to search for those with TTDOF other than GR? In order to answer this question, a class of MG theories dubbed minimally modified gravity (MMG) has been proposed in \cite{Lin:2017oow}. The term ``minimally modified'' generally indicates that we modify GR without changing its degrees of freedom (DOF's). We will use both the equivalent terms TTDOF and MMG throughout this paper.  At first glance, this search seems to have only a slim hope of success because violating the conditions in the Lovelock theorem generally increases the number of DOF's in the theory. This is exemplified by most of commonly considered MG theories. In this work, we shall abandon the 4-dimensional spacetime diffeomorphism and assume the 3-dimensional spatial diffeomorphism. Normally, new DOF's  arise when the symmetry of the theory is reduced if nothing else is done. However, reducing the symmetry also provides us with a larger space of theories, in which novel MMG theories may in fact exist. 

Besides carefully treading beyond Lovelock's assumptions, another motivation for us to search for MMG theories in the spatially covariant framework is that in comparison with other kinds of MG, e.g., scalar-tensor theories (see \cite{Langlois:2018dxi,Kobayashi:2019hrl} for reviews) (which introduce an extra scalar field as part of the gravitational interactions on top of the usual tensor field), it is simpler and more efficient to enlarge the space of theories in the spatially covariant framework, although both approaches could be related together by choosing the time slicing as the hypersurfaces of constant scalar field (namely the unitary gauge) \cite{DeFelice:2018ewo}. The main reason is that in the spatially covariant framework time and space are separated naturally, which allows one to deal with the temporal derivative and spatial derivative independently. Therefore, it is easy to avoid the Ostrogradsky ghost problem by keeping the temporal derivative up to the first order \cite{Ostrogradsky:1850fid,Woodard:2015zca} and to extend the scope of the  space of theories by pushing the spatial derivatives up to even infinite orders. On the other hand, in the scalar-tensor theories where the general covariance is manifest, the orders of time and space derivatives are always the same. As a result, the Ostrogradsky ghost(s)  will arise generally when the order of the derivatives is higher than the first order. Nevertheless, several ghost-free (namely healthy) higher order scalar-tensor (HOST) theories have  been discovered. In particular, as the generalizations of the $k$-essence theory \cite{ArmendarizPicon:1999rj,Chiba:1999ka} up to the second order in derivatives, the Horndeski/generalized galileon theories \cite{Horndeski:1974wa,Deffayet:2011gz,Kobayashi:2011nu} and the quadratic/cubic degenerate HOST (DHOST) theories \cite{Langlois:2015cwa,Langlois:2015skt,BenAchour:2016fzp} are the milestones of the HOST theories. When fixing the so-called unitary gauge with $\phi = \phi (t)$, these HOST theories can be projected onto the spatially covariant framework.  

The idea of modifying gravity by keeping only the spatial diffeomorphism can be traced back to the ghost condensation theory \cite{ArkaniHamed:2003uy} and was further developed with different purposes in the Effective Field Theory of inflation/dark energy \cite{Cheung:2007st,Gubitosi:2012hu} as well as in the Ho\v{r}ava gravity \cite{Horava:2009uw,Horava:2008ih}. In \cite{Gao:2014soa,Gao:2014fra},  a ghost-free framework for the spatially covariant gravity (SCG) theories was proposed,  which was further extended in \cite{Gao:2018znj,Gao:2019lpz} by including the velocity of the lapse function and in \cite{Gao:2018izs} with a non-dynamical scalar field. In \cite{Lin:2017oow}, a special spatially covariant framework where, by construction, the action is linear in the lapse function and where the number of DOF's is generally 2.5 was suggested as the starting point to search for MMG theories. In order to remove the half unwanted DOF, a self-consistency condition must be imposed. By solving the  consistency condition, some concrete novel MMG models were found and applied to cosmology \cite{Aoki:2018brq,Lin:2018mip,DeFelice:2020eju,DeFelice:2020cpt} and black holes \cite{DeFelice:2020onz}. Another special TTDOF theory based on the GLPV theory \cite{Gleyzes:2014dya}, where the spatial derivatives enter  the action only through the 3-dimensional extrinsic and intrinsic curvatures, was proposed in \cite{Iyonaga:2018vnu}.  Although the number of DOF's of this theory is three in general, the authors  imposed a condition to eliminate the extra DOF on the cosmological background, which yields a concrete example of TTDOF theory. Such a theory was dubbed the ``extended cuscuton''\textemdash as inspired by the first novel MMG model, the cuscuton theory \cite{Afshordi:2006ad}, of which the cosmology was discussed in \cite{Iyonaga:2020bmm}. In the general framework of SCG \cite{Gao:2014soa}, which comprises both the regions of \cite{Lin:2017oow} and \cite{Iyonaga:2018vnu}, two more general TTDOF conditions were identified in \cite{Gao:2019twq}, which include the previous conditions found in \cite{Lin:2017oow} and \cite{Iyonaga:2018vnu} as special cases. The theories satisfying these two conditions are dubbed the SCG with TTDOF. In particular, by solving the two TTDOF conditions, a novel example of MMG model was found. 

All the investigations for the MMG theories mentioned above start from the Lagrangian and derive the TTDOF conditions by performing a Hamiltonian analysis. A shortcoming of this ``Lagrangian approach'' is that the TTDOF conditions are generally complicated and difficult to solve because one first needs to make a Legendre transformation before obtaining the Hamiltonian for the constraint analysis, in which counting the number of DOF's is transparent. One way to overcome this shortcoming is to  work at the level of Hamiltonian from the beginning and to perform the Hamiltonian analysis directly. In \cite{Mukohyama:2019unx}, the authors re-constructed the MMG proposed in \cite{Lin:2017oow} by starting from the Hamiltonian and found a simpler expression for the self-consistency condition. By solving the simplified condition, a MMG theory dubbed the  $f\left(\mathcal{H}\right)$ theory was proposed. It was also shown that a cosmological model based on the $f\left(\mathcal{H}\right)$ theory (the ``kink model'')  fits the Planck data better than the $\Lambda$CDM model \cite{Aoki:2020oqc}. 

In this work, we will take an even more straightforward and aggressive approach in constructing the MMG theories. We will not only start directly from the Hamiltonian, where we assume that spatial diffeomorphism symmetry is preserved, but will also introduce an additional auxiliary constraint to assist us in locating the MMG theories in the  space of theories. As mentioned above, even without higher temporal derivatives,  there are three DOF's in the spatially covariant gravity theories in general. Thus additional constraints are necessary in order to reduce the number of DOF's. A similar idea firstly appeared in \cite{Aoki:2018zcv} (see also \cite{Lin:2018mip,Aoki:2019rvi}), where the author introduced a particular additional constraint to fix the novel gauge symmetry appearing in \cite{Lin:2017oow}. This gauge fixing condition was required in order to couple with matter consistently \cite{Aoki:2018brq}, which was also used in \cite{Aoki:2020lig,Aoki:2020iwm,Aoki:2020ila} to heal the pathology of the $D \rightarrow 4$ Einstein-Gauss-Bonnet gravity proposed in \cite{Glavan:2019inb}. 

In the current paper, we extend this idea while searching for MMG theories. Without loss of generality, we will keep the auxiliary constraint as well as the canonical Hamiltonian as two arbitrary functions from the beginning, provided that the spatial diffeomorphism is preserved. This setup is however still too lax and the number of DOF's is more than two generally. We thus need additional conditions to specify the form of the arbitrary functions in the total Hamiltonian in such a way that we are able to ``minimalize'' the theory, i.e., to obtain the subspace of MMG theories. We dub these additional conditions, which are found via a detailed Hamiltonian analysis, the ``minimalizing conditions''. As we will show, there is plenty of space of theories in which these conditions can be satisfied. 

Note that different approaches to constructing MMGs are generally complementary in their scope. Indeed, a given approach will have a tendency to point towards a specific region in the space of theories. For instance, using the construction in \cite{Aoki:2018brq}, even though the construction is rather generic, it will be difficult to produce examples from \cite{Mukohyama:2019unx}, and vice versa. For this reason, on top of the previous arguments of efficiency, it is in any case worthwhile to explore novel MMG frameworks as this will have a chance to lead to new interesting examples. Finally, note that in \cite{Aoki:2018brq} a classification of MMG theories was introduced, distinguishing type-I theories, which are equivalent to GR in vacuum, and type-II theories, their complementary set. Since in this work we only consider theories in the vacuum, we will naturally investigate the space of type-II theories. 

This paper is organized as follows. In section \ref{sec:The-Hamiltonian-with}, we will illustrate how and why we choose the total Hamiltonian (\ref{H_T}) as our starting point. In section \ref{sec:The-auxiliary-conditions}, we will find the minimalizing conditions in two cases: with and without a first-class constraint. In section \ref{sec:A-concrete-example:}, by using the minimalizing conditions, we will give an interesting example corresponding to the case without the first-class constraint and discuss the modified dispersion relation of the gravitational waves. We will finally conclude in section \ref{sec:Conclusion}.

\section{The Hamiltonian with an auxiliary constraint} \label{sec:The-Hamiltonian-with}

We will start from the total Hamiltonian of the form
\begin{eqnarray}
H_{\mathrm{T}} & = & \int \mathrm{d}^{3}x\Big[\mathscr{H}\left(N,h_{ij},\pi^{ij};\nabla_{i}\right)+N^{i}\mathcal{H}_{i}\nonumber \\
 &  & +\lambda^{i}\pi_{i}+\lambda\pi+\nu\varphi\left(N,h_{ij},\pi^{ij};\nabla_{i}\right)\Big],\label{H_T}
\end{eqnarray}
where, using the Arnowitt-Deser-Misner (ADM) formalism, $\left\{ N,N^{i},h_{ij};\pi,\pi_{i},\pi^{ij}\right\} \equiv\left\{ \Phi_{I};\Pi^{I}\right\} $
are the lapse function, shift vector, induced metric and their respective conjugate
momenta. We denote the set of canonical fields by  $\Phi_{I}$
and the set of conjugate momenta by $\Pi^{I}$ for short.  $\left\{ N^{i},\lambda^{i},\lambda,\nu\right\}$ is the set of Lagrange multipliers corresponding to the
constraints
\begin{eqnarray}
\mathcal{H}_{i} & \equiv & \pi\nabla_{i}N+\pi_{j}\nabla_{i}N^{j}+\sqrt{h}\nabla_{j} \left(\frac{\pi_{i}N^{j}}{\sqrt{h}}\right) \nonumber\\
& & -2\sqrt{h}\nabla_{j} \left(\frac{\pi_{i}^{j}}{\sqrt{h}}\right) \approx 0_{i},\label{H_i}
\end{eqnarray}
\begin{equation}
\pi_{i}\approx 0_{i},\qquad\pi\approx 0,\qquad\varphi\left(N,h_{ij},\pi^{ij};\nabla_{i}\right)\approx 0,\label{cstr}
\end{equation}
respectively.
Here, $\nabla_{i}$ is the spatially covariant derivative 
compatible with $h_{ij}$, and $\varphi$ is
an auxiliary constraint which is a free function of $\left(N,h_{ij},\pi^{ij};\nabla_{i}\right)$. Throughout this work, ``$\approx$''
represents a “weak equality” that holds only on the subspace $\Gamma_{\text{C}}$
of the phase space defined by the constraints. 
Lastly, $\mathscr{H}$ is another free function of $\left(N,h_{ij},\pi^{ij};\nabla_{i}\right)$
defining our theory. It should be emphasized that the 3-dimensional
Ricci tensor $R_{ij}$ has been included implicitly in  $\mathscr{H}\left(N,h_{ij},\pi^{ij};\nabla_{i}\right)$
and $\varphi\left(N,h_{ij},\pi^{ij};\nabla_{i}\right)$. 
The total Hamiltonian is chosen in the form (\ref{H_T}) due to the following reasons.
\begin{itemize}
\item First, in order to retain the 3-dimensional spatial
diffeomorphism symmetry, we must keep $\pi_{i}\approx0_{i}$ and $\mathcal{H}_{i}\approx0_{i}$
as the particular first-class constraints following Dirac's terminology, which generate the 3-dimensional spatial diffeomorphism, exactly as
what happens in GR. In (\ref{H_i}), we extend the definition 
of the usual momentum constraint $\mathcal{H}_{i}\approx0_{i}$ (e.g., in \cite{Gao:2014fra,Mukohyama:2015gia,Saitou:2016lvb}) in such a way
that for any quantity $Q$ that weakly vanishes, the Poisson bracket between $\mathcal{H}_{i}$ and
  $Q$ is always weakly vanishing, i.e.,
\begin{equation}
\left[\mathcal{H}_{i}\left(\vec{x}\right),Q\left(\vec{y}\right)\right]\approx0,\qquad\forall Q\approx0,\label{=00005BH_i,Q=00005D}
\end{equation}
which has been shown in \cite{Gao:2018znj}. The Poisson bracket $\left[\mathcal{F},\mathcal{G}\right]$
is defined by
\begin{eqnarray}
 \left[\mathcal{F},\mathcal{G}\right] &\equiv & \int \mathrm{d}^{3}z\sum_{I}\bigg(\frac{\delta\mathcal{F}}{\delta\Phi_{I}\left(\vec{z}\right)}\frac{\delta\mathcal{G}}{\delta\Pi^{I}\left(\vec{z}\right)} \nonumber \\
 & & \qquad \qquad -\frac{\delta\mathcal{F}}{\delta\Pi^{I}\left(\vec{z}\right)}\frac{\delta\mathcal{G}}{\delta\Phi_{I}\left(\vec{z}\right)}\bigg).\label{PB}
\end{eqnarray}
\item Second, to simplify the discussion, in this work the lapse function $N\left(t,\vec{x}\right)$
is assumed to be non-dynamical. Indeed, it has been shown in
\cite{Gao:2018znj} that a dynamical lapse leads generally to new DOF's,
thus going against the aim of this work. Although it would be very
interesting to search for MMG theories in such extended framework, we leave this for a future study.
Therefore, the conjugate momentum of the lapse is a constraint, i.e.,
$\pi\approx0$ in our theory. Generally, this is a second-class constraint
since the lapse function $N$ enters the total Hamiltonian in
a non-minimally way \cite{Mukohyama:2019unx}.
\item Last but not least, we justify the introduction of the auxiliary
constraint $\varphi\approx0$ as follows. If we do not introduce the auxiliary
constraint,  one can count
the number of DOF's of the system by calculating the so-called Dirac
matrix, of which the entries are the Poisson brackets among the different constraints. It is well known that $\mathcal{H}_{i}\approx0_{i}$
are generated from the time evolution (which is also called the consistency condition) of $\pi_{i}\approx0_{i}$ (thus as secondary constraints), if one starts from the Lagrangian aspect.
Generally, the time evolution 
of $\pi\approx 0$ will also yield a secondary constraint $\dot{\pi}\approx0$.
The time evolution of any constraint $Q$ is given by
\begin{equation}
\dot{Q}\equiv\frac{\partial Q}{\partial t}+\left[Q,H_{\mathrm{T}}\right]\approx 0,\qquad\forall Q\approx 0.\label{csc_cdt}
\end{equation}
 The consistency condition of $\dot{\pi}\approx0$ only fixes the
Lagrange multiplier $\lambda$ and does not generate further constraints
any more. Therefore, $\pi_{i}\approx0_{i}$, $\mathcal{H}_{i}\approx0_{i}$,
$\pi\approx0$ and $\dot{\pi}\approx0$ are all  the constraints
in the theory when $\varphi$ is absent. It is straightforward to check
that $\pi_{i}\approx0_{i}$ and $\mathcal{H}_{i}\approx0_{i}$ are
first-class and $\pi\approx0$ and $\dot{\pi}\approx0$ are second-class
by using (\ref{=00005BH_i,Q=00005D}) and (\ref{PB}). We thus count
the number of DOF's as
\begin{eqnarray}
\#_{\mathrm{dof}} & = & \frac{1}{2}\left(\#_{\mathrm{var}}\times2-\#_{\mathrm{1st}}\times2-\#_{\mathrm{2nd}}\right)\nonumber \\
 & = & \frac{1}{2}\left(10\times2-6\times2-2\right)=3.\label{DOF=00003D3}
\end{eqnarray}
For the sake of reducing the number of DOF's from three to two, one modifies either
the type or the number of the constraints in the system. The former
has been discussed in \cite{Mukohyama:2019unx}. In this work, we try
the latter approach, that is we introduce the auxiliary constraint $\varphi\approx0$
which is kept as general as possible, i.e., a general function of $\left(N,h_{ij},\pi^{ij};\nabla_{i}\right)$. The remaining freedom of the theory is the other free
function $\mathscr{H}\left(N,h_{ij},\pi^{ij};\nabla_{i}\right)$. 
\end{itemize}
For the above reasons, we restrict our considerations to the total Hamiltonian of the form (\ref{H_T}). Since the constraints $\mathcal{H}_{i}\approx0_{i}$
and $\pi_{i}\approx0_{i}$ are always the first-class constraints
corresponding to spatial diffeomorphism invariance, for convenience, we can
split (\ref{H_T}) into two parts
\begin{equation}
H_{\mathrm{T}}=H_{\mathrm{P}}+\int \mathrm{d}^{3}x\left(N^{i}\mathcal{H}_{i}+\lambda^{i}\pi_{i}\right), \label{H_T_split}
\end{equation} 
by denoting 
\begin{equation}
H_{\mathrm{P}}\equiv\int \mathrm{d}^{3}x\left(\mathscr{H}+\lambda\pi+\nu\varphi\right),\label{H_P}
\end{equation}
and we dub (\ref{H_P}) the ``partial'' Hamiltonian $H_{\mathrm{P}}$. The two parts in (\ref{H_T_split}) are decoupled from each other. Whether the number of DOF's can be reduced is completely encoded in $H_{\mathrm{P}}$.  We can just neglect the $\left\{ N^{i},\pi_{i}\right\} $-sector in the following
discussion and focus on $H_{\mathrm{P}}$ in order to find the MMG theories that respect the spatial diffeomorphism. In particular, in the time evolution (\ref{csc_cdt}), $H_{\mathrm{T}}$ is  replaced by $H_{\mathrm{P}}$ with an 8-dimensional phase space. One can however restore the neglected part without any difficulty in the discussion.

By using the partial Hamiltonian $H_{\mathrm{P}}$ (\ref{H_P}), the consistency
conditions of the constraints $\pi\approx0$ and $\varphi\approx0$
in (\ref{cstr}) are given by 
\begin{eqnarray}
\dot{\pi}\left(\vec{x}\right) & = & \left[\pi\left(\vec{x}\right),H_{\mathrm{P}}\right]=\int \mathrm{d}^{3}x^{\prime}\Big(\left[\pi\left(\vec{x}\right),\mathscr{H}\left(\vec{x}^{\prime}\right)\right]\nonumber \\
 & + & \nu\left[\pi\left(\vec{x}\right),\varphi\left(\vec{x}^{\prime}\right)\right]\Big)\approx0,\label{pi_dot}
\end{eqnarray}
\begin{eqnarray}
\dot{\varphi}\left(\vec{x}\right) & = & \left[\varphi\left(\vec{x}\right),H_{\mathrm{P}}\right]=\int \mathrm{d}^{3}x^{\prime}\big(\left[\varphi\left(\vec{x}\right),\mathscr{H}\left(\vec{x}^{\prime}\right)\right]\nonumber \\
 & + & \lambda\left[\varphi\left(\vec{x}\right),\pi\left(\vec{x}^{\prime}\right)\right]+\nu\left[\varphi\left(\vec{x}\right),\varphi\left(\vec{x}^{\prime}\right)\right]\Big)\approx 0.\label{vaphi_dot}
\end{eqnarray}
Since generally we have
\begin{equation}
\left[\pi\left(\vec{x}\right),\varphi\left(\vec{y}\right)\right]\neq0\quad\text{and}\quad\left[\varphi\left(\vec{x}\right),\varphi\left(\vec{y}\right)\right]\neq0,\label{=00005Bpi,=00005Cphi=00005D}
\end{equation}
 (\ref{pi_dot}) and (\ref{vaphi_dot}) fix the Lagrange multipliers
$\nu$ and $\lambda$ respectively and there  is no further constraint.
By now, we find the partial Dirac matrix as given in Table \ref{tab:The-partial-Dirac},
\begin{table}
\begin{centering}
\begin{tabular}{c|cc}
\hline 
$[\cdot(\vec{x}),\cdot(\vec{y})]$ & $\pi$ & $\varphi$\tabularnewline
\hline 
$\pi$ & 0 & $[\pi(\vec{x}),\varphi(\vec{y})]$\tabularnewline
$\varphi$ & $[\varphi(\vec{x}),\pi(\vec{y})]$ & $[\varphi(\vec{x}),\varphi(\vec{y})]$\tabularnewline
\hline 
\end{tabular}
\par\end{centering}
\caption{The partial Dirac matrix \label{tab:The-partial-Dirac}}
\end{table}
which means $\pi\approx0$ and $\varphi\approx0$ are second-class,
and the number of DOF's of the partial Hamiltonian (\ref{H_P}) (which is actually the total number of DOF's of the full theory) can be counted as
\begin{eqnarray}
\#_{\mathrm{dof}} & = & \frac{1}{2}\left(8-\#_{\mathrm{1st}}\times2-\#_{\mathrm{2nd}}\right)\nonumber \\
 & = & \frac{1}{2}\left(8-2\right)=3.
\end{eqnarray}
Comparing with (\ref{DOF=00003D3}), which corresponds to the case without the
auxiliary constraint, we can see that introducing an auxiliary constraint
in a naive way does not change the number of DOF's because the auxiliary
constraint prevents further constraints from being generated. In order to
achieve the goal of finding MMG theories, some particular conditions should be imposed
to the free functions $\mathscr{H}\left(N,h_{ij},\pi^{ij};\nabla_{i}\right)$
and $\varphi\left(N,h_{ij},\pi^{ij};\nabla_{i}\right)$ so that we
could have enough constraints to narrow the original
space of theories down to the MMG subspace. In
the next section, we will find out the TTDOF conditions,
which we will hereafter dub the ``minimalizing conditions'', by extending the Hamiltonian analysis of
(\ref{H_P}).

\section{The minimalizing conditions} \label{sec:The-auxiliary-conditions}

In order to reduce the number of DOF's of (\ref{H_P}) from three to two, i.e., to minimalize the theory, further constraints are required, coming from the degeneracy of the partial Dirac matrix shown in Table \ref{tab:The-partial-Dirac}. According to whether there is a first-class constraint, there are two parallel approaches, which we shall discuss in the following two subsections, respectively.

\subsection{Approach with a first-class constraint\label{subsec:Approach-with-first-class}}

First of all, the partial Dirac matrix is  degenerate only if the first Poisson
bracket in (\ref{=00005Bpi,=00005Cphi=00005D}) weakly vanishes, i.e.,
\begin{equation}
\left[\pi\left(\vec{x}\right),\varphi\left(\vec{y}\right)\right]=-\frac{\delta\varphi\left(\vec{y}\right)}{\delta N\left(\vec{x}\right)}\approx0.\label{phi_N}
\end{equation}
Then generally $\dot{\pi}\approx0$ (\ref{pi_dot}) becomes another
constraint and the Lagrange multiplier $\nu$ will be fixed by the
consistency condition of $\dot{\pi}\approx0$, meaning that there  is
no further constraint. 
However, normally $\pi\approx0$, $\dot{\pi}\approx0$,
and $\varphi\approx0$ are three second-class constraints and only
half of the DOF is killed, which leads to an odd-dimensional phase space
\cite{Li:2009bg,Lin:2017oow,Gao:2018znj,Gao:2019twq} that is physically
inconsistent.  In order to kill the whole DOF instead of half of it, one way is to
further require 
\begin{equation}
\left[\pi\left(\vec{x}\right),\dot{\pi}\left(\vec{y}\right)\right]=\int \mathrm{d}^{3}z\frac{\delta^{2}\mathscr{H}\left(\vec{z}\right)}{\delta N\left(\vec{x}\right)\delta N\left(\vec{y}\right)}\approx0,\label{H_0_NN}
\end{equation}
then $\pi\approx0$ will become a first-class constraint. Under the
requirements (\ref{phi_N}) and (\ref{H_0_NN}), the partial Dirac
matrix is given in Table \ref{tab:diracmatrix_firstclass},
\begin{table}
\begin{centering}
\begin{tabular}{c|ccc}
\hline 
$[\cdot(\vec{x}),\cdot(\vec{y})]$ & $\pi$ & $\varphi$ & $\dot{\pi}$\tabularnewline
\hline 
$\pi$ & 0 & $0$ & $0$\tabularnewline
$\varphi$ & 0 & $\times$ & $\times$\tabularnewline
$\dot{\pi}$ & $0$ & $\times$ & $\times$\tabularnewline
\hline 
\end{tabular}
\par\end{centering}
\caption{The partial Dirac matrix with first-class constraint}\label{tab:diracmatrix_firstclass}
\end{table}
where ``$\times$'' means that the Poisson bracket does not vanish weakly in general. In this case, the number of DOF's is counted as
\begin{eqnarray}
\#_{\mathrm{dof}} & = & \frac{1}{2}\left(8-\#_{\mathrm{1st}}\times2-\#_{\mathrm{2nd}}\right)\nonumber \\
 & = & \frac{1}{2}\left(8-1\times2-2\right)=2,
\end{eqnarray}
therefore we have the desired number of DOF's. 

In this approach, the minimalizing conditions (\ref{phi_N}) and (\ref{H_0_NN}) are required, which restrict the form of the functions $\varphi\left(N,h_{ij},\pi^{ij};\nabla_{i}\right)$
and $\mathscr{H}\left(N,h_{ij},\pi^{ij};\nabla_{i}\right)$, respectively.
An advantage of this approach is that the minimalizing conditions
(\ref{phi_N}) and (\ref{H_0_NN}) can be solved analytically. In
the strong equality case (i.e., the equality holds in the whole phase space), we solve (\ref{phi_N}) and (\ref{H_0_NN})
as
\begin{equation}
\varphi=\varphi_{0}\left(h_{ij},\pi^{ij};\nabla_{i}\right),\label{phi_0}
\end{equation}
and
\begin{equation}
\mathscr{H}=\mathcal{V}\left(h_{ij},\pi^{ij};\nabla_{i}\right)+N\mathcal{H}_{0}\left(h_{ij},\pi^{ij};\nabla_{i}\right),\label{NH_0+V}
\end{equation}
where $\varphi_{0}$, $\mathcal{H}_{0}$ and $\mathcal{V}$ are three
arbitrary functions. On the other hand, if (\ref{phi_N}) and
(\ref{H_0_NN}) only hold weakly, i.e., the right-hand-sides of (\ref{phi_N}) and (\ref{H_0_NN})
are considered to be ``linear combination''~\footnote{The coefficients in front of the constraints could be arbitrary which means they could be spatial differential operators and/or could involve the constraints as well.} of the constraints, one can show that the solutions in the weak equality
case are equivalent to the solutions in the strong equality case (\ref{phi_0})
and (\ref{NH_0+V}) up to redefinitions of the Lagrange multipliers.

To summarize, (\ref{phi_0}) and (\ref{NH_0+V}) are the general
solutions of (\ref{phi_N}) and (\ref{H_0_NN}).  In this
approach, we determine the MMG theory as 
\begin{eqnarray}
H_{\mathrm{T}} & = & \int \mathrm{d}^{3}x\big[\mathcal{V}\left(h_{ij},\pi^{ij};\nabla_{i}\right)+N\mathcal{H}_{0}\left(h_{ij},\pi^{ij};\nabla_{i}\right)\nonumber \\
 &  &+\lambda\pi +\nu\varphi_{0}\left(h_{ij},\pi^{ij};\nabla_{i}\right)+N^{i}\mathcal{H}_{i}+\lambda^{i}\pi_{i}\Big].\label{H_P_2}
\end{eqnarray}
One concrete and special example of  (\ref{H_P_2}) has
been discussed in \cite{Aoki:2020lig,Aoki:2020iwm}.

\subsection{Approach without a first-class constraint \label{subsec:Approach-I}}

There is another, parallel, choice in order to eliminate the residual half
DOF. Instead of requiring (\ref{H_0_NN}) as the companion of the necessary condition (\ref{phi_N}), it is also possible to require the second Poisson bracket in (\ref{=00005Bpi,=00005Cphi=00005D})
to be weakly vanishing 
\begin{eqnarray}
0 & \approx & \left[\varphi\left(\vec{x}\right),\varphi\left(\vec{y}\right)\right]\nonumber \\
 & = & \int d^{3}z\left(\frac{\delta\varphi\left(\vec{x}\right)}{\delta h_{mn}\left(\vec{z}\right)}\frac{\delta\varphi\left(\vec{y}\right)}{\delta\pi^{mn}\left(\vec{z}\right)}-\left(\vec{x}\leftrightarrow\vec{y}\right)\right).\label{=00005Bphi,phi=00005D}
\end{eqnarray}
As a result, generally $\dot{\varphi}\approx0$ becomes an additional constraint according to (\ref{vaphi_dot}).
It is easy to check that $\dot{\pi}\approx0$ and $\dot{\varphi}\approx0$
satisfy their consistency condition defined by (\ref{csc_cdt}), thus
the partial Dirac matrix in this approach takes the form in Table \ref{tab:diracmatrix_nofirstclass}.
\begin{table}
\begin{centering}
\begin{tabular}{c|cccc}
\hline 
$[\cdot(\vec{x}),\cdot(\vec{y})]$ & $\pi$ & $\varphi$ & $\dot{\pi}$ & $\dot{\varphi}$\tabularnewline
\hline 
$\pi$ & 0 & $0$ & $\times$ & $\times$\tabularnewline
$\varphi$ & 0 & 0 & $\times$ & $\times$\tabularnewline
$\dot{\pi}$ & $\times$ & $\times$ & $\times$ & $\times$\tabularnewline
$\dot{\varphi}$ & $\times$ & $\times$ & $\times$ & $\times$\tabularnewline
\hline 
\end{tabular}
\par\end{centering}
\caption{The partial Dirac matrix without first-class constraint. }\label{tab:diracmatrix_nofirstclass}
\end{table}
Therefore, the number of DOF's corresponding to this system is 
\begin{eqnarray}
\#_{\mathrm{dof}} & = & \frac{1}{2}\left(8-\#_{\mathrm{1st}}\times2-\#_{\mathrm{2nd}}\right)\nonumber \\
 & = & \frac{1}{2}\left(8-4\right)=2,
\end{eqnarray}
which is exactly what we are looking for. In other words, if the auxiliary
constraint $\varphi\approx0$ satisfies the minimalizing conditions
(\ref{phi_N}) and (\ref{=00005Bphi,phi=00005D}) simultaneously,
the number of DOF's of the theory (\ref{H_T}) is exactly two. 

As we can see, through this approach, only the form of function $\varphi\left(N,h_{ij},\pi^{ij};\nabla_{i}\right)$
is restricted by (\ref{phi_N}) and (\ref{=00005Bphi,phi=00005D})
and the function $\mathscr{H}\left(N,h_{ij},\pi^{ij};\nabla_{i}\right)$
remains totally free. One simple and special solution of the minimalizing
conditions (\ref{phi_N}) and (\ref{=00005Bphi,phi=00005D}) is 
\begin{equation}
\varphi=\tilde{\varphi}\left(h_{ij},\pi^{ij}\right).
\end{equation}
Thus the theory described by the Hamiltonian
\begin{eqnarray}
H_{\mathrm{T}} & = & \int \mathrm{d}^{3}x\big[\mathscr{H}\left(N,h_{ij},\pi^{ij};\nabla_{i}\right)+\lambda\pi\nonumber \\
 &  & +\nu\tilde{\varphi}\left(h_{ij},\pi^{ij}\right)+N^{i}\mathcal{H}_{i}+\lambda^{i}\pi_{i}\Big],\label{H_T-1}
\end{eqnarray}
 contains TTDOF only, where $\mathscr{H}\left(N,h_{ij},\pi^{ij};\nabla_{i}\right)$
and $\tilde{\varphi}\left(h_{ij},\pi^{ij}\right)$ are two free functions.
As an implementation of this approach, we will give a more interesting
example in  section \ref{sec:A-concrete-example:}.

To summarize the above, once the free functions $\varphi\left(N,h_{ij},\pi^{ij};\nabla_{i}\right)$
and $\mathscr{H}\left(N,h_{ij},\pi^{ij};\nabla_{i}\right)$ simultaneously
satisfy one of the two sets of the minimalizing conditions (\ref{phi_N})
and (\ref{H_0_NN}) or (\ref{phi_N}) and (\ref{=00005Bphi,phi=00005D}),
the theory (\ref{H_T}) is an MMG theory.

\section{A concrete example} \label{sec:A-concrete-example:}

\subsection{The Cayley-Hamilton construction with a linear auxiliary constraint}

In this section, we are going to give an interesting example corresponding
to the case without a first-class constraint discussed in subsection
\ref{subsec:Approach-I}. 

First, we check in the appendix \ref{sec:The-linear-ansatz}
that the following linear ansatz for the auxiliary constraint $\varphi$
\begin{equation}
\hat{\varphi}=c_{1}\left(t\right)\pi_{i}^{i}+c_{2}\left(t\right)\sqrt{h}R_{i}^{i}+c_{3}\left(t\right)\sqrt{h}\nabla^{2}\left(\frac{\pi_{i}^{i}}{\sqrt{h}}\right),\label{lin_ansatz}
\end{equation}
is a solution of the auxiliary conditions (\ref{phi_N}) and (\ref{=00005Bphi,phi=00005D}), where  the coefficients $c_{1}\left(t\right)$, $c_{2}\left(t\right)$ and $c_{3}\left(t\right)$ are arbitrary functions of time. Second, for simplicity, we restrict $\mathscr{H}$ to an arbitrary function of $\left(N,h_{ij},\pi^{ij},R_{ij}\right)$, i.e., 
\begin{equation}
\mathscr{H}=\mathscr{H}\left(N,h_{ij},\pi^{ij},R_{ij}\right).\label{scr_H}
\end{equation}
However, under the restriction of (\ref{scr_H}), according to the generalized Cayley-Hamilton theorem \cite{Mertzios:1986cht} (see appendix \ref{sec:The-generalized-Cayley-Hamilton} for details) we know that the number of the independent scalars (i.e., traces) constructed from $\pi^{ij}$ and $R_{ij}$ (using $h_{ij}$ to raise or to lower indices) is limited, which can be chosen as follows
\begin{eqnarray}
 &  & \Big\{ R_{i}^{i},\,R_{j}^{i}R_{i}^{j},\,R_{j}^{i}R_{k}^{j}R_{i}^{k};\,\pi_{i}^{i},\,\pi_{j}^{i}\pi_{i}^{j},\,\pi_{j}^{i}\pi_{k}^{j}\pi_{i}^{k};\nonumber \\
 &  & \quad R_{j}^{i}\pi_{i}^{j},\,R_{j}^{i}R_{k}^{j}\pi_{i}^{k},\,R_{j}^{i}\pi_{k}^{j}\pi_{i}^{k}\Big\}.\label{9_traces}
\end{eqnarray}
Therefore, the free function $\mathscr{H}$ in (\ref{scr_H}) can be recast equivalently to
\begin{equation}
\mathscr{H}=\mathscr{H}^{\left(\text{C.H.}\right)}\left(N,\mathscr{R}^{A},\varPi^{A},\mathscr{Q}^{A}\right),\label{H^C.H.}
\end{equation}
where we denote
\begin{equation}
\mathscr{R}^{A}\equiv\left\{ R_{i}^{i},\,R_{j}^{i}R_{i}^{j},\,R_{j}^{i}R_{k}^{j}R_{i}^{k}\right\},
\end{equation}
\begin{equation}
\varPi^{A}\equiv\left\{ \pi_{i}^{i},\,\pi_{j}^{i}\pi_{i}^{j},\,\pi_{j}^{i}\pi_{k}^{j}\pi_{i}^{k}\right\},
\end{equation}
and
\begin{equation}
\mathscr{Q}^{A}\equiv\left\{ R_{j}^{i}\pi_{i}^{j},\,R_{j}^{i}R_{k}^{j}\pi_{i}^{k},\,R_{j}^{i}\pi_{k}^{j}\pi_{i}^{k}\right\},
\end{equation}
with $A$  from 1 to 3. To summarize, we determine the following total Hamiltonian 
\begin{eqnarray}
& & H_{\mathrm{T}}^{\left(\text{C.H.}\right)}
 =  \int \mathrm{d}^{3}x\bigg[\mathscr{H}^{\left(\text{C.H.}\right)}+N^{i}\mathcal{H}_{i}+\lambda^{i}\pi_{i}+\lambda\pi \nonumber \\
 &  & +\nu\left(c_{1}(t)\pi_{i}^{i}+c_{2}(t)\sqrt{h}R_{i}^{i}+c_{3}(t)\sqrt{h}\nabla^{2}\left(\frac{\pi_{i}^{i}}{\sqrt{h}}\right)\right)\bigg],\quad \label{H_T_CH}
\end{eqnarray}
which satisfies the auxiliary conditions (\ref{phi_N}) and (\ref{=00005Bphi,phi=00005D}) and propagates only TTDOF. Since it is determined using the Cayley-Hamilton theorem, we dub the construction in (\ref{H_T_CH}) the Cayley-Hamilton construction with a linear auxiliary constraint.

For simplicity we may consider a subset of the general Cayley-Hamilton construction (\ref{H_T_CH}) by restricting the free function $\mathscr{H}$ to the following form~\footnote{Since $\pi^{ij}$ and $R_{ij}$ are two symmetric matrices, the more general $\mathscr{H}$ given in (\ref{H^C.H.}) can be written in terms of the six traces $(\mathscr{R}^{A},\varPi^{A})$ for configurations that simultaneously diagonalize $\pi_{i}^{j}$ and $R_{i}^{j}$. However, variations around such a configuration do not necessarily diagonalize $\pi_{i}^{j}$ and $R_{i}^{j}$ simultaneously and thus one needs to include $\mathscr{Q}^{A}$ as well in general. In other words, (\ref{6-traces}) is nothing but a simplifying ansatz.},
\begin{equation}
\mathscr{H}=\mathscr{H}^{\left(\text{C.H.}\right)}\left(N,\mathscr{R}^{A},\varPi^{A}\right),\label{6-traces}
\end{equation}
which depends only on the lapse $N$ and the six traces $(\mathscr{R}^{A},\varPi^{A})$ as the independent variables and thus does not depend on the mixed traces $\mathscr{Q}^{A}$. We dub (\ref{H_T_CH}) with the subset (\ref{6-traces}) the unmixed Cayley-Hamilton construction with a linear auxiliary constraint. In the next subsection we study this subset.

\subsection{The dispersion relation}
\label{subsec:dispersionrelation-gw}

For the sake of illustrating the properties of the Cayley-Hamilton construction (\ref{H_T_CH}) in a cosmological setting, we will derive the dispersion relation of gravitational waves in this model. For simplicity, however, we restrict our consideration to the subset (\ref{6-traces}). First we obtain the action corresponding to the Hamiltonian (\ref{H_T_CH}) with (\ref{6-traces}) by performing a Legendre transformation
\begin{eqnarray}
& & S^{\left(\text{C.H.}\right)}  =  -\int \mathrm{d}t\mathrm{d}^{3}x\bigg[\mathscr{H}^{\left(\text{C.H.}\right)}-2N\pi^{ij}K_{ij} \nonumber \\
 &  & +\nu\left(c_{1}(t)\pi_{i}^{i}+c_{2}(t)\sqrt{h}R_{i}^{i}+c_{3}(t)\sqrt{h}\nabla^{2}\left(\frac{\pi_{i}^{i}}{\sqrt{h}}\right)\right)\bigg],\quad \label{S^C.H.}
\end{eqnarray}
where $\pi^{ij}$ should be understood as the solution of
\begin{equation}
2NK_{ij}-\left(c_{1}\nu+c_{3}\nabla^{2}\nu\right)h_{ij}=\frac{\partial\mathscr{H}^{\left(\text{C.H.}\right)}}{\partial\pi^{ij}},\label{pi^ij_eq}
\end{equation}
which depends on the concrete form of $\mathscr{H}^{\left(\text{C.H.}\right)}$. Since we keep it as a general function in this work, we are only able to solve (\ref{pi^ij_eq}) perturbatively. 

Next, we consider tensor perturbations around a flat FLRW background as
\begin{equation}
 \nu=\bar{\nu}\left(t\right),\quad N=1,\quad N^{i}=0, \quad 
 h_{ij}=a\left(t\right)^{2}\mathfrak{g}_{ij},
\end{equation}
corresponding to the four-dimensional metric 
\begin{equation}
\mathrm{d}s^{2}=-\mathrm{d}t^{2}+a\left(t\right)^{2}\mathfrak{g}_{ij}\mathrm{d}x^{i}\mathrm{d}x^{j},
\end{equation}
where
\begin{equation}
\mathfrak{g}_{ij}\equiv\delta_{ij}+\gamma_{ij}+\frac{1}{2!}\gamma_{ik}\gamma^{k}{}_{j}+\frac{1}{3!}\gamma_{ik}\gamma^{k}{}_{l}\gamma^{l}{}_{j}+\cdots,
\end{equation}
with the tensor perturbation $\gamma^{i}{}_{j}$ satisfying the transverse and traceless conditions
\begin{equation}
\partial_{i}\gamma^{i}{}_{j}=0,\qquad\gamma^{i}{}_{i}=0.
\end{equation}
(We have turned off the scalar- and vector-type perturbations.) Note that in this subsection, spatial indices are raised and lowered by $\delta^{ij}$ and $\delta_{ij}$. 

By solving (\ref{pi^ij_eq}) for $\pi^{ij}$ order by order and substituting it back into the action (\ref{S^C.H.}) with (\ref{6-traces}), we have the following quadratic
action
\begin{eqnarray}
 &  & S_{2}^{\left(\text{C.H.}\right)}=\int dtd^{3}x\frac{1}{4}\Big(\mathcal{G}_{0}\left(t\right)\dot{\gamma}_{ij}\dot{\gamma}^{ij} \nonumber \\
 &  & +\mathcal{W}_{0}\left(t\right)\gamma_{ij}\frac{\Delta}{a^{2}}\gamma^{ij}-\mathcal{W}_{2}\left(t\right)\gamma_{ij}\frac{\Delta^{2}}{a^{4}}\gamma^{ij}\Big),\label{2nd_action}
\end{eqnarray}
with
\begin{eqnarray}
\mathcal{G}_{0}\left(t\right) & = & \mp\bigg[\left.\frac{\partial\mathscr{H}^{\left(\text{C.H.}\right)}}{\partial\varPi^{2}}\right|_{\left(0\right)}^{2}-3\left.\frac{\partial\mathscr{H}^{\left(\text{C.H.}\right)}}{\partial\varPi^{3}}\right|_{\left(0\right)}\nonumber \\
 & \times & \left(\left.\frac{\partial\mathscr{H}^{\left(\text{C.H.}\right)}}{\partial\varPi^{1}}\right|_{\left(0\right)}-2H\left(t\right)+c_{1}\bar{\nu}\left(t\right)\right)\bigg]^{-\frac{1}{2}},\label{G_0}
\end{eqnarray}
\begin{equation}
\mathcal{W}_{0}\left(t\right)=-\left(\left.\frac{\partial\mathscr{H}^{\left(\text{C.H.}\right)}}{\partial\mathscr{R}^{1}}\right|_{\left(0\right)}+a^{3}c_{2}\bar{\nu}\left(t\right)\right),
\end{equation}
and
\begin{equation}
\mathcal{W}_{2}\left(t\right)=\left.\frac{\partial\mathscr{H}^{\left(\text{C.H.}\right)}}{\partial\mathscr{R}^{2}}\right|_{\left(0\right)}, \label{W_2}
\end{equation}
where ``$\left.\right|_{\left(0\right)}$'' denotes taking the values on the background and $H\left(t\right)\equiv\dot{a}/a$ is the Hubble parameter. The plus sign in (\ref{G_0}) should be taken when $\left.\frac{\partial\mathscr{H}^{\left(\text{C.H.}\right)}}{\partial\varPi^{2}}\right|_{\left(0\right)}>0$ and the minus sign when $\left.\frac{\partial\mathscr{H}^{\left(\text{C.H.}\right)}}{\partial\varPi^{2}}\right|_{\left(0\right)}<0$. Therefore, one needs to impose
\begin{equation}
 \left.\frac{\partial\mathscr{H}^{\left(\text{C.H.}\right)}}{\partial\varPi^{2}}\right|_{\left(0\right)}>0
\end{equation}
to prevent tensor perturbations from becoming ghosts. From (\ref{2nd_action}), we can immediately read the dispersion relation \cite{Gao:2019liu} as follows
\begin{eqnarray}
\frac{\omega_{\text{T}}^{2}}{a^{2}} & = & \frac{\mathcal{W}_{0}\left(\tau\right)}{\mathcal{G}_{0}\left(\tau\right)}\frac{k^{2}}{a^{2}}+\frac{\mathcal{W}_{2}\left(\tau\right)}{\mathcal{G}_{0}\left(\tau\right)}\frac{k^{4}}{a^{4}} \nonumber \\
 & = & \frac{k^{2}}{a^{2}}\mathcal{G}_{0}^{-1}\Big[\left.\frac{\partial\mathscr{H}^{\left(\text{C.H.}\right)}}{\partial\mathscr{R}^{2}}\right|_{\left(0\right)}\frac{k^{2}}{a^{2}} \nonumber \\
 &  & -\left.\frac{\partial\mathscr{H}^{\left(\text{C.H.}\right)}}{\partial\mathscr{R}^{1}}\right|_{\left(0\right)}-a^{3}c_{2}\bar{\nu}\left(t\right)\Big].\label{dispersion_rlt}
\end{eqnarray}
On large scales, the speed of gravitational waves $c_{\text{T}}=\omega_{\text{T}}/k=1$
when
\begin{equation}
-\left.\frac{\partial\mathscr{H}^{\left(\text{C.H.}\right)}}{\partial\mathscr{R}^{1}}\right|_{\left(0\right)}=\mathcal{G}_{0}+a^{3}c_{2}\bar{\nu}\left(t\right).\label{c_T=00003D1}
\end{equation}
In light of the constraint from the observation of the speed of gravitational
waves \cite{Monitor:2017mdv,TheLIGOScientific:2017qsa}, we should impose the following constraint
\begin{equation}
-3\times10^{-15}<\frac{\mathcal{W}_{0}}{\mathcal{G}_{0}}-1<7\times10^{-16}.\label{w0/g0-1}
\end{equation}
On the other hand, according to \cite{Abbott:2017vtc}, the modified
dispersion relation has the following upper bound
\begin{equation}
\left|\frac{\mathcal{W}_{2}}{\mathcal{G}_{0}}\right|<10^{-19}\,\text{peV}^{-2},\label{W2/G0}
\end{equation}
where $1\,\,\text{peV}\simeq h\times250\,\,\text{Hz}$ with $h$ the Planck
constant.

\section{Conclusion} \label{sec:Conclusion}

In this work, we have searched for minimally modified gravity (MMG) theories\textemdash those carrying the same two tensorial degrees of freedom (TTDOF) as GR\textemdash within the spatially covariant framework and focusing on the Hamiltonian formalism. We have first specified a fairly general and practical total Hamiltonian (\ref{H_T}) as our starting point, in which the spatial diffeomorphism is preserved and the lapse function is set to be non-dynamical so that the initial DOF's are the two tensor and one scalar modes.  In order to eliminate the unwanted scalar  DOF, i.e., to restrict the initial space of theories to the MMG subspace, we have introduced an auxiliary constraint $\varphi\approx0$ (\ref{cstr}), which is the key assumption of the current work.  This auxiliary constraint is kept as general as possible, i.e., as an arbitrary function $\varphi\left(N,h_{ij},\pi^{ij};\nabla_{i}\right)$ from the beginning. However, the number of DOF's is still three with a general $\varphi$. Hence, we need additional conditions (\ref{phi_N})  together with (\ref{H_0_NN}) or (\ref{=00005Bphi,phi=00005D}), which are dubbed the ``minimalizing conditions'', to narrow the space of theories down to the MMG subspace. These conditions are found by performing the constraint analysis of the total Hamiltonian (\ref{H_T}). Depending on whether a first-class constraint exists or not, we find two different approaches to our purpose.

In the first approach\textemdash the one with a first-class constraint\textemdash discussed in subsection \ref{subsec:Approach-with-first-class}, we have derived the minimalizing conditions (\ref{phi_N}) and (\ref{H_0_NN}) which restrict the form of the free functions $\varphi\left(N,h_{ij},\pi^{ij};\nabla_{i}\right)$ and $\mathscr{H}\left(N,h_{ij},\pi^{ij};\nabla_{i}\right)$ to (\ref{phi_0}) and (\ref{NH_0+V}), respectively, up to redefinition of Lagrange multipliers. In this way, the total Hamiltonian (\ref{H_T}) is restricted to (\ref{H_P_2}), which represents a large class of MMG theories where the lapse function appears as a Lagrange multiplier. 

In the second approach\textemdash the one without a first-class constraint\textemdash discussed in subsection \ref{subsec:Approach-I}, two minimalizing conditions (\ref{phi_N}) and (\ref{=00005Bphi,phi=00005D}) have also been derived. In this approach, only the form of the auxiliary constraint $\varphi\left(N,h_{ij},\pi^{ij};\nabla_{i}\right)$ is restricted, and the function $\mathscr{H}\left(N,h_{ij},\pi^{ij};\nabla_{i}\right)$ remains totally free. One special solution is given in (\ref{H_T-1}), where the auxiliary constraint does not involve any spatial derivative. In section \ref{sec:A-concrete-example:}, a more interesting solution (\ref{H_T_CH}) with a linear ansatz for the auxiliary constraint (\ref{lin_ansatz}) has been constructed, where the free function $\mathscr{H}^{\left(\text{C.H.}\right)}\left(N,\mathscr{R}^{A},\varPi^{A}\right)$ (\ref{H^C.H.}) is determined using the generalized Cayley-Hamilton theorem. As a simple application, restricting our consideration to the subset (\ref{6-traces}) for simplicity, we have studied tensor perturbations in the corresponding action (\ref{S^C.H.}) up to the quadratic order on a FLRW background, and derived the modified dispersion relation (\ref{dispersion_rlt}) for the gravitational waves. We have found that on large scales, the speed of gravitational waves is unity when (\ref{c_T=00003D1}) is satisfied. In general the free function $\mathscr{H}^{\left(\text{C.H.}\right)}\left(N,\mathscr{R}^{A},\varPi^{A}\right)$ is constrained by (\ref{w0/g0-1}) and (\ref{W2/G0}) (with (\ref{G_0})-(\ref{W_2})) from the observations.

A few comments are in order. First, in this work we have introduced only a single auxiliary constraint to illustrate the idea and also to simplify the discussion, but multiple auxiliary constraints are allowed in principle and may render it easier to construct MMG theories. Second, we have worked in the spatially covariant framework where the lapse function is assumed to be non-dynamical. On the other hand, a dynamical lapse function is allowed \cite{Gao:2018znj}, although this by itself will lead to one extra scalar mode in general. It is thus interesting to search for the MMG subspace in the spatially covariant framework with a dynamical lapse function, which may require more additional constraints/conditions in order to reduce the number of DOF's from four to two. Third, note that since we have exclusively worked in the vacuum, those theories that we have investigated fall naturally into the space of type-II theories, according to the classification introduced in \cite{Aoki:2018brq}, aside from trivial  cases. Indeed, type-II theories differ from type-I ones in that they are not equivalent to GR in the vacuum as is evident from the nonlinear dispersion relation of gravitational waves studied in subsection~\ref{subsec:dispersionrelation-gw}. The study of the coupling to matter, besides the question of its consistency within the above Hamiltonian analysis, is an interesting question since non-minimal couplings may introduce further richness and variety in the theories obtained. Fourth, since we got two types of MMG theories with two different approaches according to whether there is a first-class constraint or not, it may be worthwhile to ask what  the relation between these two classes of MMG theories is, and what  the equivalent MMG theories are after getting rid of the auxiliary field in the Lagrangians. Moreover, we may also examine if the minimalizing conditions derived in this work are related to any (gauge) symmetry in the MMG theories. Recently, the symmetries for scalarless scalar-tensor theories (dubbed the scalarless symmetry) were discussed in \cite{Tasinato:2020fni}, which may be  useful for the discussion of the symmetries of MMG theories. Last but not least, we may find more interesting concrete examples, which could then be investigated more deeply in the cosmological context. We will come back to these questions in the near future.

\acknowledgments

X.G. was partly supported by the Natural Science Foundation of China (NSFC) under the grant No. 11975020. The work of S.M.\ was supported in part by Japan Society for the Promotion of Science Grants-in-Aid for Scientific Research No.~17H02890, No.~17H06359, and by World Premier International Research Center Initiative, MEXT, Japan.

\appendix

\section{The linear ansatz $\hat{\varphi}$}\label{sec:The-linear-ansatz}

In this appendix we show that the linear ansatz (\ref{lin_ansatz})
\begin{equation}
\hat{\varphi}=c_{1}\left(t\right)\pi_{i}^{i}+c_{2}\left(t\right)\sqrt{h}R_{i}^{i}+c_{3}\left(t\right)\sqrt{h}\nabla^{2}\left(\frac{\pi_{i}^{i}}{\sqrt{h}}\right),
\end{equation}
satisfies the auxiliary conditions (\ref{phi_N}) and (\ref{=00005Bphi,phi=00005D}).
First, the condition (\ref{phi_N}) is trivially satisfied.  For the condition (\ref{=00005Bphi,phi=00005D}),
we have
\begin{widetext}
\begin{eqnarray}
 &  & \int \mathrm{d}^{3}z\left(\frac{\delta\overline{\text{\ensuremath{\hat{\varphi}}}}\left[\alpha\right]}{\delta h_{mn}\left(\vec{z}\right)}\frac{\delta\overline{\text{\ensuremath{\hat{\varphi}}}}\left[\beta\right]}{\delta\pi^{mn}\left(\vec{z}\right)}-\left(\alpha\leftrightarrow\beta\right)\right)\nonumber \\
 & = & \int \mathrm{d}^{3}z\Big\{\Big[\alpha\left(c_{1}\pi^{mn}+c_{2}\sqrt{h}\left(\frac{1}{2}Rh^{mn}-R^{mn}\right)+c_{3}\frac{1}{2}h^{mn}\sqrt{h}\nabla^{k}\nabla_{k}\frac{\pi_{i}^{i}}{\sqrt{h}}\right)\nonumber \\
 &  & +c_{2}\sqrt{h}\left(\nabla^{m}\nabla^{n}\alpha-h^{mn}\nabla_{k}\nabla^{k}\alpha\right)+c_{3}\left(\nabla^{k}\nabla_{k}\alpha\right)\left(\pi^{mn}-\frac{1}{2}\pi_{i}^{i}h^{mn}\right)\Big]\nonumber \\
 &  & \times\left(c_{1}\left(t\right)\beta\left(\vec{z}\right)+c_{3}\left(t\right)\nabla^{2}\beta\left(\vec{z}\right)\right)h_{mn}-\left(\alpha\leftrightarrow\beta\right)\Big\}\nonumber \\
 & = & \int \mathrm{d}^{3}z\left[c_{2}\sqrt{h}\left(\nabla^{m}\nabla^{n}\alpha-h^{mn}\nabla_{k}\nabla^{k}\alpha\right)+c_{3}\left(\nabla^{k}\nabla_{k}\alpha\right)\left(\pi^{mn}-\frac{1}{2}\pi_{i}^{i}h^{mn}\right)\right]\nonumber \\
 &  & \times\left(c_{1}\left(t\right)\beta\left(\vec{z}\right)+c_{3}\left(t\right)\nabla^{2}\beta\left(\vec{z}\right)\right)h_{mn}-\left(\alpha\leftrightarrow\beta\right)\Big\}=0,
\end{eqnarray}
\end{widetext}
where we denote 
\begin{equation}
\overline{\text{\ensuremath{\hat{\varphi}}}}\left[\alpha\right]\equiv\int \mathrm{d}^{3}x\hat{\varphi}\left(\vec{x}\right)\alpha\left(\vec{x}\right)
\end{equation}
for short, and the variations are calculated as
\begin{equation}
\frac{\delta\overline{\text{\ensuremath{\hat{\varphi}}}}\left[\beta\right]}{\delta\pi^{mn}\left(\vec{z}\right)}=\left(c_{1}\left(t\right)\beta\left(\vec{z}\right)+c_{3}\left(t\right)\nabla^{2}\beta\left(\vec{z}\right)\right)h_{mn},
\end{equation}
and
\begin{eqnarray}
\frac{\delta\overline{\text{\ensuremath{\hat{\varphi}}}}\left[\alpha\right]}{\delta h_{mn}\left(\vec{z}\right)} & = & \alpha\Big[c_{1}\pi^{mn}+c_{2}\sqrt{h}\left(\frac{1}{2}Rh^{mn}-R^{mn}\right)\nonumber \\
 &  & +c_{3}\frac{1}{2}h^{mn}\sqrt{h}\nabla^{k}\nabla_{k}\frac{\pi_{i}^{i}}{\sqrt{h}}\Big]\nonumber \\
 &  & +c_{2}\sqrt{h}\left(\nabla^{m}\nabla^{n}\alpha-h^{mn}\nabla_{k}\nabla^{k}\alpha\right)\nonumber \\
 &  & +c_{3}\left(\nabla^{k}\nabla_{k}\alpha\right)\left(\pi^{mn}-\frac{1}{2}\pi_{i}^{i}h^{mn}\right).
\end{eqnarray}

\section{The generalized Cayley-Hamilton theorem\label{sec:The-generalized-Cayley-Hamilton}}

The usual Cayley-Hamilton theorem states that an arbitrary $n$-dimensional
matrix $\boldsymbol{A}$ over a commutative ring satisfies its own
characteristic equation
\begin{equation}
p\left(\boldsymbol{A}\right)=\sum_{i=0}^{n}c_{i}\boldsymbol{A}^{i}=\boldsymbol{0},\label{p(A)}
\end{equation}
where the characteristic equation is defined by the characteristic
polynomial of $\boldsymbol{A}$, 
\begin{equation}
p\left(\lambda\right)=\det\left(\lambda\boldsymbol{I}-\boldsymbol{A}\right).
\end{equation}
Here $\bm{A}^{0}\equiv \bm{I}_{n}$ is  the $n$-dimensional identity
matrix. The coefficients $c_{i}$ in (\ref{p(A)}) can be calculated
by the following determinants
\begin{equation}
c_{n-m}=\frac{\left(-1\right)^{m}}{m!}\left|\begin{array}{ccccc}
\text{tr}\boldsymbol{A} & m-1 & 0 & \cdots\\
\text{tr}\boldsymbol{A}^{2} & \text{tr}\boldsymbol{A} & m-2 & \cdots\\
\vdots & \vdots &  &  & \vdots\\
\text{tr}\boldsymbol{A}^{m-1} & \text{tr}\boldsymbol{A}^{m-2} & \cdots & \cdots & 1\\
\text{tr}\boldsymbol{A}^{m} & \text{tr}\boldsymbol{A}^{m-1} & \cdots & \cdots & \text{tr}\boldsymbol{A}
\end{array}\right|,
\end{equation}
with $n\leq m\leq0$. Especially, we have 
\begin{equation}
c_{n}=1,\qquad c_{0}\equiv\left(-1\right)^{n}\det\left(\boldsymbol{A}\right).
\end{equation}
For instance, in the 3-dimensional $(n=3)$ case, we have
\begin{eqnarray}
p\left(\boldsymbol{A}\right) & = & \boldsymbol{A}^{3}-\left(\begin{array}{c}
\text{tr}\boldsymbol{A}\end{array}\right)\boldsymbol{A}^{2}+\frac{1}{2}\left[\left(\text{tr}\boldsymbol{A}\right)^{2}-\text{tr}\left(\boldsymbol{A}^{2}\right)\right]\boldsymbol{A}\nonumber \\
 &  & -\det\left(\boldsymbol{A}\right)\boldsymbol{I}_{3}=\boldsymbol{0},
\end{eqnarray}
where $\boldsymbol{I}_{3}$ is the 3-dimensional identity matrix and
the determinant of $\boldsymbol{A}$ can be expressed by
\begin{equation}
\det\left(\boldsymbol{A}\right)=\frac{1}{6}\left[\left(\text{tr}\boldsymbol{A}\right)^{3}-3\text{tr}\left(\boldsymbol{A}^{2}\right)\left(\text{tr}\boldsymbol{A}\right)+2\text{tr}\left(\boldsymbol{A}^{3}\right)\right].
\end{equation}
From the above, we know that any higher matrix power $\boldsymbol{A}^{k}$
($k\geq n$) can be written as a matrix polynomial of degree at most
$n-1$ and by taking the trace of the characteristic equation (\ref{p(A)}),
we also know that any trace of the higher matrix power $\text{tr}\left(\boldsymbol{A}^{l}\right)$
($l\geq n+1$) can be expressed by the traces polynomial of degree
at most $n$. Therefore the independent traces in the $n=3$ case can be
chosen as 
\begin{equation}
\left\{ \text{tr}\boldsymbol{A},\,\text{tr}\left(\boldsymbol{A}^{2}\right),\,\text{tr}\left(\boldsymbol{A}^{3}\right)\right\} .
\end{equation}

The generalization of Cayley-Hamilton theorem with two $n\times n$
matrices $\boldsymbol{A}$ and $\boldsymbol{B}$ was discussed in
\cite{Mertzios:1986cht}. In this case, the
characteristic equation (\ref{p(A)}) is generalized to the following
set of characteristic equations
\begin{eqnarray}
f_{i}\left(\boldsymbol{A},\boldsymbol{B}\right) & = & \left\langle \boldsymbol{A}^{i},\boldsymbol{B}^{n-i}\right\rangle \nonumber \\
 & + & \underset{_{\left(k,l\right)\neq\left(0,0\right)}}{\sum_{l=0}^{i}\sum_{k=l}^{n-i+l}}q_{kl}\left\langle \boldsymbol{A}^{i-l},\boldsymbol{B}^{n-i+l-k}\right\rangle =\boldsymbol{0},\label{f_i}
\end{eqnarray}
where 
\begin{equation}
\left\langle \boldsymbol{A}^{i},\boldsymbol{B}^{n-i}\right\rangle \equiv\boldsymbol{A}^{i}\boldsymbol{B}^{n-i}+\boldsymbol{A}^{i-1}\boldsymbol{B}^{n-i}\boldsymbol{A}+\cdots\boldsymbol{B}^{n-i}\boldsymbol{A}^{i},
\end{equation}
with $0\leq i\leq n$. The algorithm for the coefficients $q_{kl}$,
$\left\{ k=1,2,\cdots,n\,\,\text{and}\,\,l\leq k\right\} $ can be
found by the Faddeev-LeVerrier's algorithm for singular
systems
\begin{equation}
q_{kl}=-\frac{1}{k}\text{tr}\left[\boldsymbol{B}\boldsymbol{R}_{k-1,l}+\boldsymbol{A}\boldsymbol{R}_{k-1,l-1}\right],
\end{equation}
where
\begin{equation}
\boldsymbol{R}_{kl}=\boldsymbol{B}\boldsymbol{R}_{k-1,l}+\boldsymbol{A}\boldsymbol{R}_{k-1,l-1}+q_{kl}\boldsymbol{I}_{n},
\end{equation}
with the boundary conditions
\begin{equation}
\boldsymbol{R}_{00}=\boldsymbol{I}_{n},
\end{equation}
\begin{equation}
\boldsymbol{R}_{-k,l}=\boldsymbol{R}_{k,-l}=\boldsymbol{0},\qquad\text{for}\,\,k\geq1,\,\,l\geq1,
\end{equation}
and
\begin{equation}
\boldsymbol{R}_{kl}=\boldsymbol{0},\qquad\text{for}\,\,l>k\geq0.
\end{equation}
For instance, in the 3-dimensional ($n=3$) case, (\ref{f_i}) are listed as
follows
\begin{equation}
f_{0}\left(\boldsymbol{A},\boldsymbol{B}\right)=\boldsymbol{B}^{3}+q_{10}\boldsymbol{B}^{2}+q_{20}\boldsymbol{B}+q_{30}\boldsymbol{I}_{3}=\boldsymbol{0},
\end{equation}
\begin{eqnarray}
f_{1}\left(\boldsymbol{A},\boldsymbol{B}\right) & = & \boldsymbol{A}\boldsymbol{B}^{2}+\boldsymbol{B}\boldsymbol{A}\boldsymbol{B}+\boldsymbol{B}^{2}\boldsymbol{A}+q_{10}\left(\boldsymbol{A}\boldsymbol{B}+\boldsymbol{B}\boldsymbol{A}\right)\nonumber \\
 & + & q_{11}\boldsymbol{B}^{2}+q_{20}\boldsymbol{A}+q_{21}\boldsymbol{B}+q_{31}\boldsymbol{I}_{3}=\boldsymbol{0},
\end{eqnarray}
\begin{eqnarray}
f_{2}\left(\boldsymbol{A},\boldsymbol{B}\right) & = & \boldsymbol{A}^{2}\boldsymbol{B}+\boldsymbol{A}\boldsymbol{B}\boldsymbol{A}+\boldsymbol{B}\boldsymbol{A}^{2}+q_{11}\left(\boldsymbol{A}\boldsymbol{B}+\boldsymbol{B}\boldsymbol{A}\right)\nonumber \\
 & + & q_{10}\boldsymbol{A}^{2}+q_{21}\boldsymbol{A}+q_{22}\boldsymbol{B}+q_{32}\boldsymbol{I}_{3}=\boldsymbol{0},
\end{eqnarray}
\begin{equation}
f_{3}\left(\boldsymbol{A},\boldsymbol{B}\right)=\boldsymbol{A}^{3}+q_{11}\boldsymbol{A}^{2}+q_{22}\boldsymbol{A}+q_{33}\boldsymbol{I}_{3}=\boldsymbol{0},
\end{equation}
where the coefficients are 
\begin{equation}
q_{10}=-\text{tr}\boldsymbol{B},\qquad q_{11}=-\text{tr}\boldsymbol{A},
\end{equation}
\begin{equation}
q_{20}=\frac{1}{2}\left[\left(\text{tr}\boldsymbol{B}\right)^{2}-\text{tr}\left(\boldsymbol{B}^{2}\right)\right],
\end{equation}
\begin{equation}
q_{21}=\text{tr}\boldsymbol{A}\text{tr}\boldsymbol{B}-\text{tr}\left(\boldsymbol{A}\boldsymbol{B}\right),\quad q_{22}=\frac{1}{2}\left[\left(\text{tr}\boldsymbol{A}\right)^{2}-\text{tr}\left(\boldsymbol{A}^{2}\right)\right],
\end{equation}
\begin{equation}
q_{30}=-\frac{1}{6}\left[\left(\text{tr}\boldsymbol{B}\right)^{3}-3\text{tr}\left(\boldsymbol{B}^{2}\right)\left(\text{tr}\boldsymbol{B}\right)+2\text{tr}\left(\boldsymbol{B}^{3}\right)\right],
\end{equation}
\begin{eqnarray}
q_{31} & = & -\frac{1}{2}\Big[2\text{tr}\left(\boldsymbol{A}\boldsymbol{B}^{2}\right)-\text{tr}\boldsymbol{A}\text{tr}\left(\boldsymbol{B}^{2}\right)\nonumber \\
 &  & -2\text{tr}\left(\boldsymbol{A}\boldsymbol{B}\right)\text{tr}\boldsymbol{B}+\text{tr}\boldsymbol{A}\left(\text{tr}\boldsymbol{B}\right)^{2}\Big],
\end{eqnarray}
\begin{eqnarray}
q_{32} & = & -\frac{1}{2}\Big[2\text{tr}\left(\boldsymbol{A}^{2}\boldsymbol{B}\right)-2\text{tr}\left(\boldsymbol{A}\boldsymbol{B}\right)\text{tr}\boldsymbol{A}\nonumber \\
 &  & +\left(\text{tr}\boldsymbol{A}\right)^{2}\text{tr}\boldsymbol{B}-\text{tr}\left(\boldsymbol{A}^{2}\right)\text{tr}\boldsymbol{B},\Big]
\end{eqnarray}
and
\begin{equation}
q_{33}=-\frac{1}{6}\left[\left(\text{tr}\boldsymbol{A}\right)^{3}-3\text{tr}\left(\boldsymbol{A}^{2}\right)\text{tr}\boldsymbol{A}+2\text{tr}\left(\boldsymbol{A}^{3}\right)\right].
\end{equation}
Similarly, the independent traces constructed by $\boldsymbol{A}$
and $\boldsymbol{B}$ could be chosen as
\begin{eqnarray}
 &  & \Big\{\text{tr}\boldsymbol{A},\text{tr}\left(\boldsymbol{A}^{2}\right),\text{tr}\left(\boldsymbol{A}^{3}\right),\text{tr}\boldsymbol{B},\text{tr}\left(\boldsymbol{B}^{2}\right),\text{tr}\left(\boldsymbol{B}^{3}\right),\nonumber \\
 &  & \text{tr}\left(\boldsymbol{A}\boldsymbol{B}\right),\text{tr}\left(\boldsymbol{A}^{2}\boldsymbol{B}\right),\text{tr}\left(\boldsymbol{A}\boldsymbol{B}^{2}\right)\Big\}.\label{trA_trB}
\end{eqnarray}


\bibliography{MMG_auxiliary_constraint}

\end{document}